\documentclass[conference]{IEEEtran}

\usepackage{color, xcolor, float, lscape, enumerate, graphicx, url, tabularx, multirow, xspace, hyperref}
\usepackage{booktabs}
\usepackage{etoolbox}
\usepackage{multirow}
\usepackage{amsfonts}
\usepackage[linesnumbered,ruled]{algorithm2e}

\usepackage{subfigure}
\usepackage{comment}
\usepackage{amsmath,amssymb,amsfonts}
\usepackage{graphicx}
\usepackage{times}
\usepackage[ruled,linesnumbered]{algorithm2e}
\usepackage{cite}
\usepackage{multicol}
\usepackage{multirow}
\usepackage{xspace}
\usepackage{xcolor}
\usepackage{listings}
\usepackage{balance}
\usepackage{multirow}
\usepackage{pdfpages}

\newcommand{\BfPara}[1]{{\noindent\bf#1.}\xspace}

\newcommand{\etal}{{\em et al.}\xspace}
\newcommand{\eg}{{\em e.g.,}\xspace}

\newcommand{\etc}{{\em etc.}\xspace}

\renewcommand{\baselinestretch}{0.905}

\definecolor{dkgreen}{rgb}{0,0.6,0}
\definecolor{gray}{rgb}{0.5,0.5,0.5}
\definecolor{mauve}{rgb}{0.58,0,0.82}
\usepackage{verbatim}
\hypersetup{  
	pdfpagemode=pagewidth,
	plainpages=false, 
	colorlinks,  
	urlcolor=red!50!black,   
	linkcolor=blue!50!black,    
	citecolor=blue!50!black,  
	bookmarksnumbered
}

\begin{document}

\title{Examining Adversarial Learning against Graph-based IoT Malware Detection Systems} \vspace{-2mm}



\author{\IEEEauthorblockN{Ahmed Abusnaina, Aminollah Khormali, Hisham Alasmary, Jeman Park, \\ Afsah Anwar, Ulku Meteriz, and Aziz Mohaisen}
\IEEEauthorblockA{ Department of Computer Science, University of Central Florida }
\IEEEauthorblockA{ \{ahmed.abusnaina, aminkhormali, hisham, parkjeman, afsahanwar, meteriz\}@knights.ucf.edu, mohaisen@cs.ucf.edu  } 
}\vspace{-5mm}

\maketitle

\begin{abstract}
The main goal of this study is to investigate the robustness of graph-based Deep Learning (DL) models used for Internet of Things (IoT) malware classification against Adversarial Learning (AL). We designed two approaches to craft adversarial IoT software, including Off-the-Shelf Adversarial Attack (OSAA) methods, using six different AL attack approaches, and Graph Embedding and Augmentation (GEA). The GEA approach aims to preserve the functionality and practicality of the generated adversarial sample through a careful embedding of a benign sample to a malicious one. Our evaluations demonstrate that OSAAs are able to achieve a misclassification rate (MR) of 100\%. Moreover, we observed that the GEA approach is able to misclassify all IoT malware samples as benign. 
\end{abstract}


\vspace{-2mm}

\section{Introduction}\label{sec:introduction}

\vspace{-1mm}

Internet of Things (IoT) devices, including sensors, voice assistants, automation tools, \etc~\cite{Gerber17}, are widely used, increasing the attack surface of the Internet due to their evolving and often insecure software. Thus, it is essential to understand IoT software to address security issues through analysis and detection~\cite{Gerber17}. However, the research work on IoT software analysis has been very limited not only in the size of the analyzed samples, but also the utilized approaches~\cite{azmoodehDCK18}. A promising direction leverages a graph-theoretic approach to analyze IoT malware. Representative static characteristics of IoT applications can be extracted from the Control Flow Graph (CFG), which can be utilized to build an automatic IoT malware detection system~\cite{AlasmaryA0CNM18}.

Machine Learning (ML) algorithms, specifically DL networks, are actively used in a wide range of applications, such as health-care, industry, cyber-security, and \etc~\cite{MohaisenAM15, khormaliA16}. However, it has been shown that ML/DL networks are vulnerable to AL, where an adversary can force the model to his desired output, \eg misclassification. Although it is an active research area, there is very little research work done on understanding the impact of AL on DL-based IoT malware detection system and practical implications~\cite{GrossePMBM17}, particularly those that utilize CFG features for detection. 

\BfPara{Goal of this study}\label{sec:contribution}
Motivated by the aforementioned issues, our main goal is generating {\em adversarial IoT software samples that (1) fool the classifier and (2) function as intended.} 

\BfPara{Approach} To tackle the above objectives, we designed two approaches to craft adversarial examples, including OSAA and GEA approaches. The OSAA approach incorporates six well-known adversarial learning methods to force the model to misclassification. Whereas, the GEA approach aims to preserve the functionality and practicality of the generated adversarial samples through a careful connection of benign graph to a malicious one.

\BfPara{Contributions} Our contributions are as follows: 1) We examined the robustness of CFG-based deep learning IoT malware detection system using two different approaches, including off-the-shelf adversarial learning algorithms and graph embedding and augmentation, while maintaining the practicality and functionality of the crafted AEs. 2) We found that the first approach can generate AEs with MR of 100\%. However, they do not guarantee the practicality and functionality of the crafted AEs, unlike the GEA approach.

\vspace{-2mm}
\section{Generating Adversarial Examples}\label{sec:methodology}
\vspace{-2mm}
In order to generate realistic AEs that preserve the functionality and practicality of the original samples we design two approaches: generic adversarial machine learning attacks and GEA. More information regarding the proposed approaches are presented in~\textsection\ref{sec:GAA} and~\textsection\ref{sec:attack_GEA}.   
\vspace{-2mm}
\subsection{Off-the-Shelf Adversarial Attacks (OSAA)} \label{sec:GAA}
\vspace{-1mm}
This approach incorporates well-established adversarial machine learning attack methods into IoT malware detection. These methods apply small perturbation into the feature space to generate AEs that lead to misclassification.

\vspace{-2mm}
\subsection{Graph Embedding and Augmentation (GEA)} \label{sec:attack_GEA}
\vspace{-1mm}

\begin{figure*}[t]
\centering
\begin{subfigure}[Original sample's CFG\label{fig:GC_Xorg}]{\includegraphics[width=0.25\textwidth]{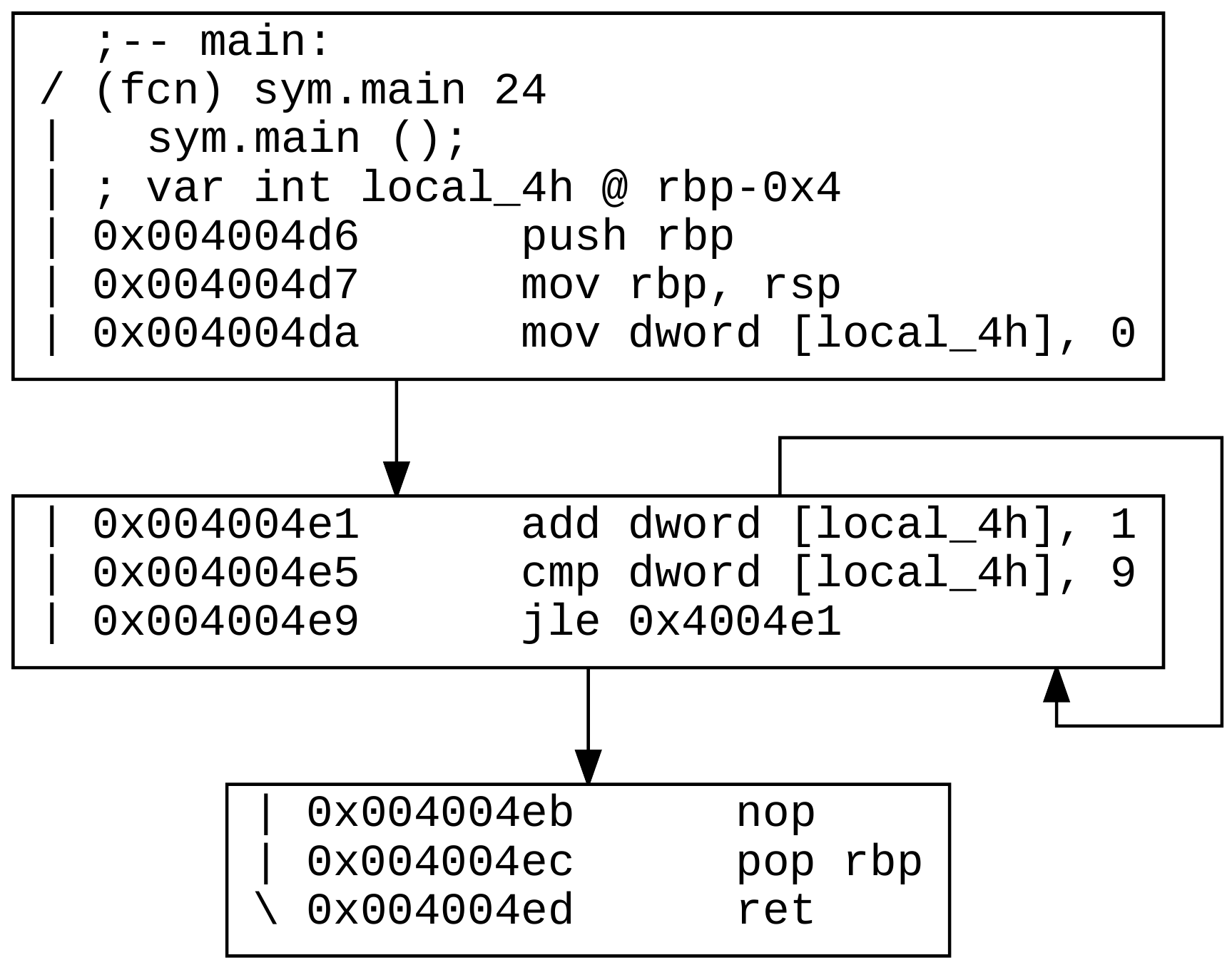}} 
\end{subfigure}
\begin{subfigure}[Taraget sample's CFG\label{fig:GC_Xsel}]{\includegraphics[width=0.27\textwidth]{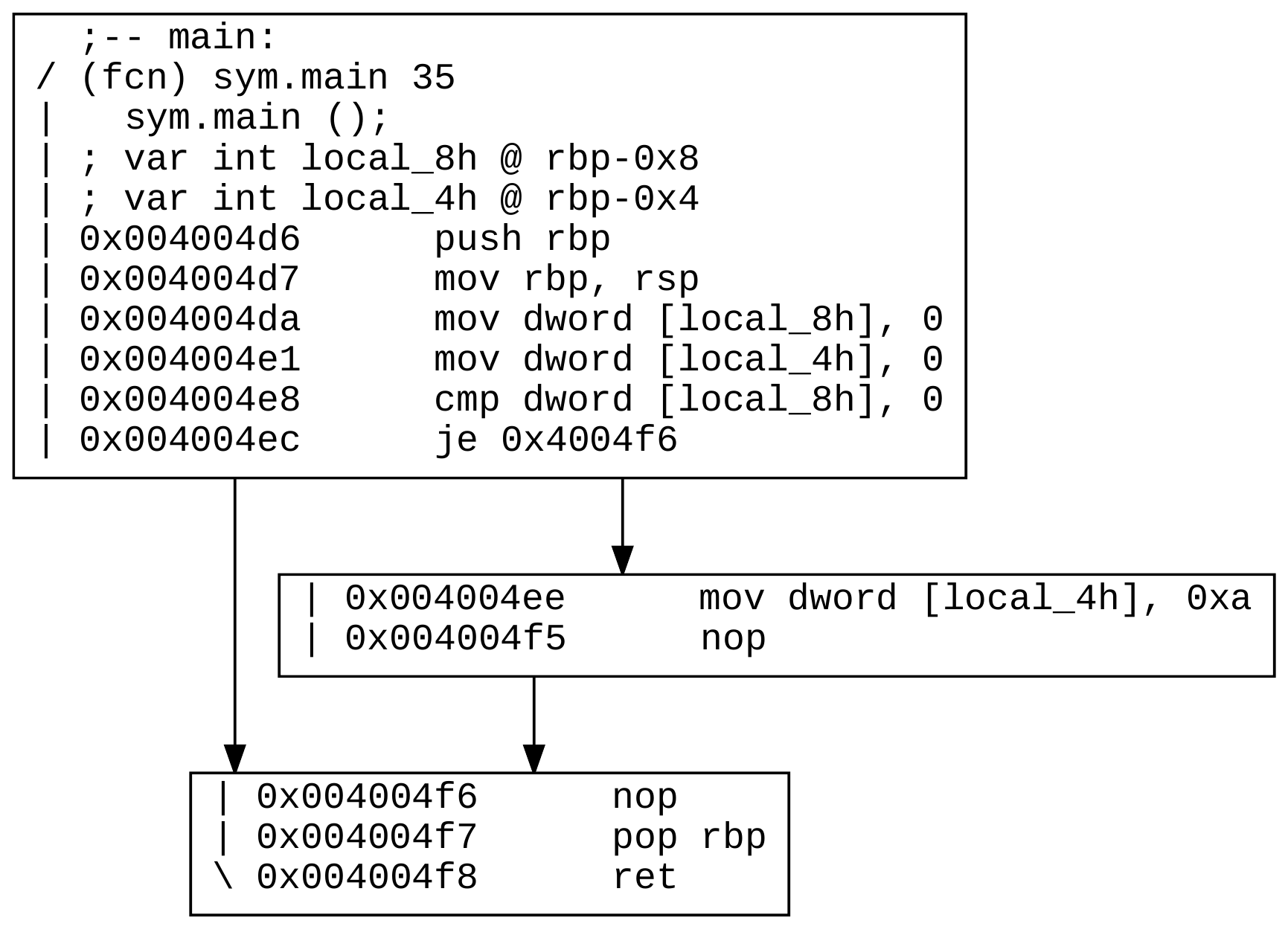}}
\end{subfigure}
\begin{subfigure}[Crafted adversarial CFG using GEA\label{fig:GC_Xcomp}]{\includegraphics[width=0.33\textwidth]{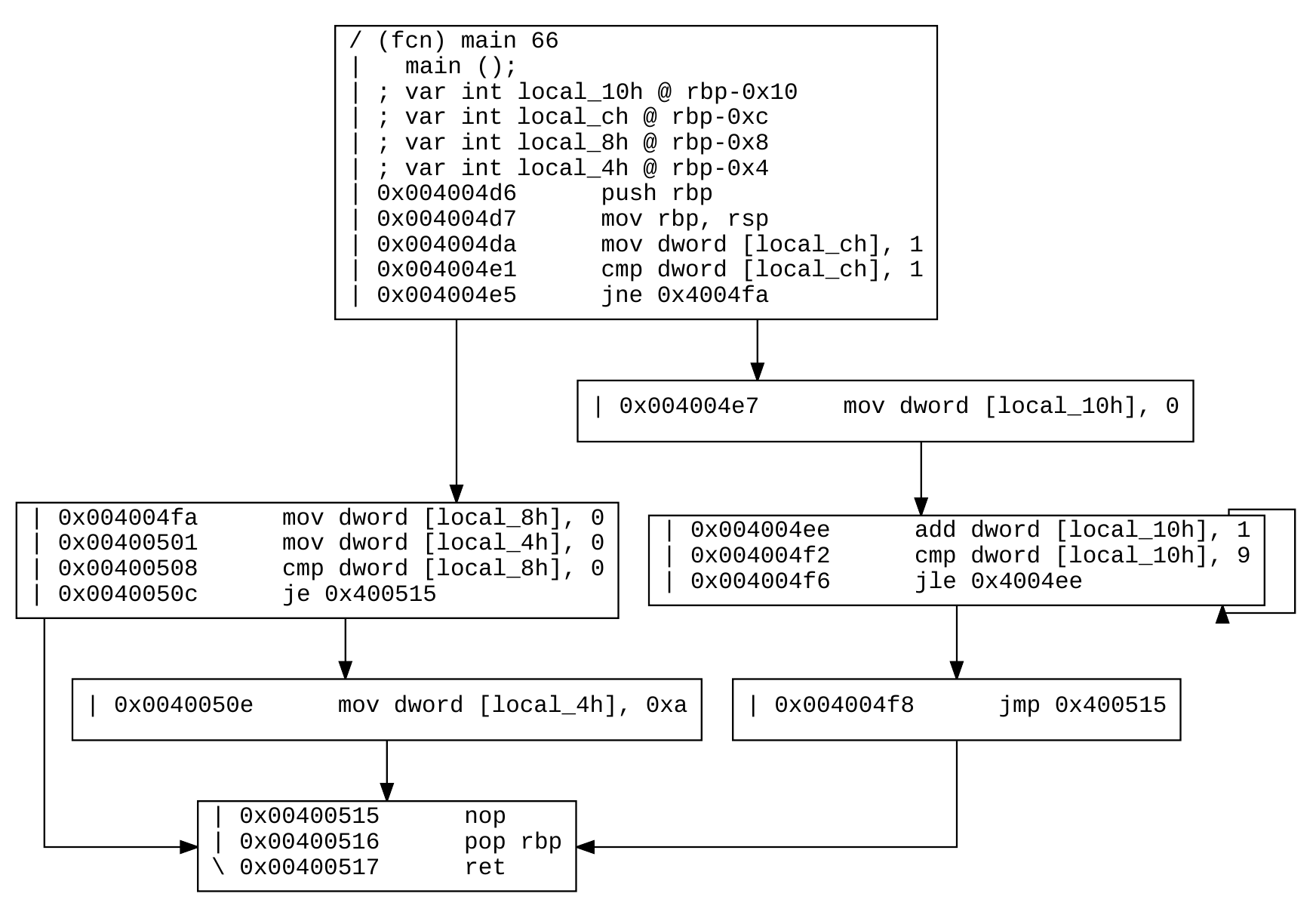}}
\end{subfigure}
\vspace{-2mm}
\caption{ A practical implementation of the GEA approach. Fig.~\ref{fig:GC_Xorg} shows the generated CFG for the original sample and used for extracting graph-based features (graph size, centralities, etc.) for graph/program classification and malware detection. \autoref{fig:GC_Xsel} shows the graph for the selected target sample generated as in Fig.~\ref{fig:GC_Xorg}. Finally, The generated adversarial graph using GEA approach. Note that this graph is obtained logically by embedding the graph in Fig.~\ref{fig:GC_Xsel} into the graph in Fig.~\ref{fig:GC_Xorg}.} 
\vspace{-6mm}

\label{fig:dynamic}
\end{figure*}

Assume an original sample $x_{org}$ and a selected target sample $x_{sel}$, our main goal is to combine the two samples while preserving the functionality and practicality of $x_{org}$ and achieving misclassification. Prior to generating the CFG for these algorithms, we compile the code using GNU Compiler Collection (GCC) command. Afterwards, Radare2 is used to extract the CFG from the binaries. \autoref{fig:GC_Xorg} and \autoref{fig:GC_Xsel} show the generated graphs for $x_{org}$ and $x_{sel}$, respectively.

\vspace{-2mm}
\section{Evaluation and Discussion}\label{sec:results}

\vspace{-2mm}
\subsection{Dataset} \label{sec:Dataset} 
\vspace{-2mm}

We obtained the CFG dataset of the IoT malware from Alasmary~\etal\cite{AlasmaryA0CNM18} to assess our proposed approach. The dataset consists of 2,281 malicious and 276 benign IoT samples. We extracted 23 different features in seven different groups, including betweenness centrality, closeness centrality, degree centrality, shortest path, density, \# of edges, and \# of nodes.  

\vspace{-3mm}
\subsection{Results \& Discussion} \label{sec:Results}
\vspace{-2mm}

\subsubsection{Deep Learning-based IoT Malware Detection System}\label{sec:Res_DLMDS}
We designed a CNN-based classifier, which distinguishes IoT malware samples from benign ones, trained over 23 CFG-based features categorized in seven groups, including betweenness centrality, closeness centrality, degree centrality, shortest path, density, \# of edges, and \# of nodes, extracted from CFGs of  2,281 malware and 276 benign samples. We achieved an accuracy rate of 97.13\% with a False Negative Rate (FNR) of 11.26\% and False Positive Rate (FPR) of 1.55\%. It is worth mentioning that the high value of FNR is due to the imbalanced number of malware and benign samples.

\subsubsection{OSAA}\label{sec:Res_GAMLA}
We implemented six generic adversarial learning attack methods to generate AE by perturbing the feature space. Overall, those approaches have shown, in general, a good performance (see~\autoref{GAttacks}). 

\begin{table}[t]
\centering
\caption{Evaluation of the generic adversarial learning attack methods. MR: misclassification rate, Avg.FG: average number of changed features,and CT: computation time.}\vspace{-2mm}
\label{GAttacks}
\scalebox{0.82}{
\begin{tabular}{c|c|c|c}
\toprule
Attack Method & MR (\%) & Avg.FG & CT (ms)\\

\hline

C\&W~\cite{Carlini017} &  100 & 12.60 & 25.30  \\ 
DeepFool~\cite{Moosavi-Dezfooli16} &  86.39 & 14.90 & 2.56  \\ 
ElasticNet~\cite{ChenSZYH18} &  100 & 5.42 & 114.18  \\ 
JSMA~\cite{PapernotMJFCS16} &  99.80 & 4.00 & 0.78  \\ 
MIM~\cite{DONGLPSZHL18}&  100 & 20.60 & 0.90  \\ 
PGD~\cite{MadryMSTV17} &  100 & 22.56 & 2.40  \\ 
\bottomrule
\end{tabular}
}\vspace{-4mm}
\end{table}

\subsubsection{GEA}\label{sec:Res_GC}
This approach is designed to generate a practical AE that fools the classifier, while preserving the functionality and practicality of the original sample. Here, we discuss the inherent overhead of the GEA approach. We investigate the impact of the size of the graph, determined by the number of the nodes in a graph, and graph density, determined by the number of edges in a graph while the number of nodes is fixed. Note that all generated samples maintain the practicality and the functionality of the original sample. The obtained results are discussed in more detail in the following.

\BfPara{Graph Size Impact}\label{sec:exp1}
We selected three graphs, as targets, from each of the benign and malicious IoT software, consisting of a minimum, median and maximum graph size, and the goal was to understand the impact of size on MR with GEA. The results are shown in~\autoref{GCNodesMB}. We found that the MR increases when the number of nodes increases, which is perhaps natural. In addition, the time needed to craft the AE is proportional to the size of the selected sample. We achieved a malware to benign MR of as high as 100\%, and a benign to malware MR of 88.04\%, while insuring that the original samples are executed as intended, a property not guaranteed with the off-the-shelf adversarial attack methods. 

\begin{table}[t]
\centering
\caption{GEA: Malware to benign (Mal2Ben) and benign to malware (Ben2Mal) misclassification rate. MR: misclassification rate, CT: computational time.}\vspace{-2mm}
\label{GCNodesMB}
\scalebox{0.82}{
\begin{tabular}{c|c|c|c|c}
\toprule
 & Size & \# Nodes & MR (\%) & CT (ms)\\

\hline

\multirow{3}{*}{Mal2Ben} & Minimum &  2 & 7.67 & 33.69  \\ 
& Median &  24 & 95.48 & 37.79  \\ 
& Maximum &  455 & 100 & 1,123.12 \\ \hline

\multirow{3}{*}{Ben2Mal} & Minimum &  1 & 30.65 & 40.65  \\ 
& Median &  64 & 57.60 & 69.23  \\ 
& Maximum &  367 & 88.04 & 473.91 \\ 

\bottomrule
\end{tabular}
}\vspace{-2mm}
\end{table}

\section{Conclusion}\label{sec:conclusion}
\vspace{-2mm}
In this work, we generated the CFGs of the IoT samples, we then extracted 23 representative features from the CFGs to train our DL model. The focus of this study is to investigate the robustness of the trained DL model. Thus, we designed two approaches, including OSAA methods and GEA. OSAA methods incorporates six different attacks to generate the AE. In our evaluation, we obtain a MR of up to 100\% using these attacks. GEA approach focuses on preserving the functionality and practicality of the generated samples, which is not guaranteed in OSAA methods. Our evaluation showed that GEA is able to misclassify all malware samples as benign. 

\vspace{-1mm}
\balance

\section{Acknowledgment} \vspace{-2mm}
This work was supported by NRF-2016K1A1A2912757, NVIDIA GPU Grant Program, Office of Research and Commercialization Fellowship, and a collaborative seed grant from the Florida Cybersecurity Center (FC2).

\footnotesize
\bibliographystyle{IEEEtran}
\bibliography{ref,conf}


\end{document}